\documentclass[10pt, a4paper]{article}
\pdfoutput=1

\usepackage[utf8]{inputenc}
\usepackage[T1]{fontenc}
\usepackage{lmodern, textcomp}
\usepackage[british]{babel}
\usepackage{csquotes}

\usepackage[heightrounded, top=4cm, bottom=4cm, left=3.5cm, right=3.5cm]{geometry}

\usepackage{eurosym, xspace}
\usepackage[nottoc]{tocbibind}
\usepackage{graphicx, subfig, float}
\usepackage[multiple]{footmisc}
\usepackage{authblk}

\usepackage[usenames, dvipsnames]{xcolor}

\usepackage[backend=bibtex8, citestyle=numeric-comp, maxbibnames=99,
			sorting=none, sortcites=true, giveninits=true]{biblatex}
\input{arxiv.def}

\usepackage{amsmath, amsfonts, amssymb}
\usepackage{cancel}

\usepackage{maths_min}

\usepackage[pdftex, breaklinks=true, linktocpage,
	colorlinks=true, urlcolor=blue, linkcolor=blue, citecolor=red,
	pdfauthor={Harold Erbin}
]{hyperref}
\usepackage[all]{hypcap}

\graphicspath{{images/}}

\numberwithin{equation}{section}

\DeclareUnicodeCharacter{00A0}{~}
\DeclareUnicodeCharacter{202F}{~}

\newcommand{\email}[1]{\thanks{\href{mailto:#1}{\nolinkurl{#1}}}}

\usepackage[sort&compress, english]{cleveref}

\hypersetup{
	pdftitle={Minisuperspace computation of the Mabuchi spectrum},
	pdfkeywords={LPTENS}
}

\title{Minisuperspace computation of the Mabuchi spectrum}

\author[1]{Corinne de Lacroix\email{lacroix@lpt.ens.fr}}
\author[1]{Harold Erbin\email{erbin@lpt.ens.fr}}
\author[2]{Eirik E. Svanes\email{esvanes@lpthe.jussieu.fr}}

\affil[1]{\textsc{Lpt}, Département de physique de l'\textsc{Ens}, École normale supérieure, \textsc{Upmc} Univ. Paris 06, \textsc{Cnrs}, \textsc{Psl} Research University, 75005 Paris, France}
\affil[1]{Sorbonne Universités, \textsc{Upmc} Univ. Paris 06, École normale supérieure, \textsc{Cnrs}, \textsc{Lpt}, 75005 Paris, France}

\affil[2]{Sorbonne Universités, \textsc{Cnrs}, \textsc{Lpthe}, \textsc{Upmc} Paris 06, \textsc{Umr} 7589, 75005 Paris, France}
\affil[2]{Sorbonne Universités, Institut Lagrange de Paris, 98 bis Bd Arago, 75014 Paris, France}

\begin{document}
\maketitle

\begin{abstract}
It was shown recently that, beside the traditional Liouville action, other functionals appear in the gravitational action of two-dimensional quantum gravity in the conformal gauge, the most important one being the Mabuchi functional.
In a letter we proposed a minisuperspace action for this theory and used it to perform its canonical quantization.
We found that the Hamiltonian of the Mabuchi theory is equal to the one of the Liouville theory and thus that the spectrum and correlation functions match in this approximation.
In this paper we provide motivations to support our conjecture.
\end{abstract}

\newpage

\hrule
\pdfbookmark[1]{\contentsname}{toc}
\tableofcontents
\bigskip
\hrule

\section{Introduction}

Two-dimensional quantum gravity is an important toy model for its four-dimensional cousin because many computations can be carried out exactly, in particular, quantum corrections due to the interaction between matter and gravity.
Hence it is essential to precisely characterize the properties of this theory in its most possible general form.

The theory is especially simple when one considers only conformal matter.
It was shown by Polyakov that the effective action, in this case, is given by the Liouville action~\cite{Polyakov:1981:QuantumGeometryBosonic}.
A huge work has been provided for defining this theory, in particular for understanding the critical exponents~\cite{David:1988:ConformalFieldTheories, Distler:1989:ConformalFieldTheory}, the spectrum~\cite{Curtright:1982:ConformallyInvariantQuantization, DHoker:1982:ClassicalQuantalLiouville, Braaten:1983:ExactOperatorSolution, Gervais:1985:Nonstandard2DCritical, Seiberg:1990:NotesQuantumLiouville} and the correlation functions~\cite{Knizhnik:1988:FractalStructure2dQuantum, Dorn:1994:TwoThreepointFunctions, Zamolodchikov:1996:StructureConstantsConformal}.
This culminated in the conformal bootstrap of Liouville theory which demonstrated that it defines a consistent CFT for any complex central charge using a non-Lagrangian description~\cite{Ribault:2015:LiouvilleTheoryCentral} (see also~\cite{Teschner:2001:LiouvilleTheoryRevisited}).

Despite the fact that four-dimensional gravity is not scale invariant (and even less conformal invariant), the coupling of two-dimensional gravity to non-conformal matter has been mostly ignored in the literature.\footnotemark{}
\footnotetext{%
    Some properties of classical gravity coupled to massive matter have been studied in~\cite{deLacroix:2016:ShortNoteDynamics}.
}%
The case of gravity coupled to a CFT perturbed by primary operators has been studied through the DDK ansatz (see for example~\cite{Seiberg:1990:NotesQuantumLiouville, Zamolodchikov:2002:ScalingLeeYangModel, Zamolodchikov:2005:PerturbedConformalField, Zamolodchikov:2006:MassiveMajoranaFermion}), but there are reasons to believe that this approach does not fit in the usual framework.
Moreover, nothing has been attempted until recently for genuine non-conformal field theories where the perturbation is not a primary operator (such as the mass term of a scalar field).
In this case, it was shown that other functionals contribute to the gravitational action, in~\cite{Ferrari:2011:RandomGeometryQuantum, Ferrari:2012:GravitationalActionsTwo, Bilal:2017:2DQuantumGravity, Bilal:2017:2DGravitationalMabuchi} for the massive scalar field (possibly with non-minimal coupling) and in~\cite{Ferrari:2014:FQHECurvedBackgrounds} for a scalar field with a linear term.

The most prominent functional that arises in the gravitational action of non-conformal matter is the Mabuchi action (also called K-energy in the mathematics literature)~\cite{Mabuchi:1986:KenergyMapsIntegrating} which modifies the dynamics of the Liouville field.
For this reason, the Mabuchi action is an essential element of a general two-dimensional quantum gravity and it is important to understand its physical properties by following the same program as for the Liouville theory -- first for the pure Mabuchi theory, and then for the coupled Liouville--Mabuchi theory.
The case of the pure Mabuchi theory is not only of academic interest since it is possible to tune the matter in order to obtain a pure Mabuchi gravity with a cosmological constant (at leading order)~\cite[sec.~3.3]{Bilal:2014:2DQuantumGravity}.
A first step has been taken in~\cite{Bilal:2014:2DQuantumGravity} where the $1$-loop string susceptibilities in the Liouville--Mabuchi and pure Mabuchi theories have been computed.

The next natural stage is to find the spectrum of the pure Mabuchi theory~\cite[sec.~3.3]{Bilal:2014:2DQuantumGravity}.
In order to tackle this problem, we rely on the minisuperspace approximation where the quantum field theory is reduced to the quantum mechanics of a point particle.
In general, the dynamics of the zero-mode is sufficient to build the Hilbert space which is made from normalizable wave functions.
We find that, in this limit and under our assumptions, the Mabuchi Hamiltonian coincides with the Liouville Hamiltonian.
As a consequence, the spectra are identical in both theories.
Using the results for the spectrum it is straightforward to provide an expression for the $3$-point function in the semi-classical limit.
These results were already reported in the companion letter~\cite{deLacroix:2016:MabuchiSpectrumMinisuperspace} and the current paper provides additional details for the justification of the minisuperspace action.
Indeed a rigorous derivation of the minisuperspace action and of the associated Hamiltonian requires a variable-area action, but the latter is not known for the Mabuchi theory.
We provide several independent derivations relying on different assumptions of the minisuperspace action for the Mabuchi action: the fact that the results agree in all cases gives a strong support to our proposal.
Finally, we stress that this computation considers the Mabuchi action in isolation as it is defined in~\cite[sec.~3.3]{Bilal:2014:2DQuantumGravity}, and in particular without the area-dependent coupling constant arising from the matter, the reason being that this piece is not universal.
Modifying the Lagrangian (for example by coupling with the Liouville action) removes some of the problems we encounter but this also answers a different question than the one we are proposing to address in this paper.
Hence we leave these studies for future works.

The paper is organized as follows.
In \cref{sec:quantum} we recall the formulation of the two-dimensional quantum gravity in the conformal gauge and we describe the properties of the Liouville and Mabuchi actions in order to set the stage for the derivation of the minisuperspace action.
We end this section by discussing the subtleties associated to the flat topology.
Then in \cref{sec:minisuperspace} we comment on the difficulties in computing rigorously the minisuperspace action and we give several derivations to motivate our proposal.
Finally, \cref{sec:quantization} contains our main result on the spectrum of the Mabuchi theory which is obtained by performing a canonical quantization of the minisuperspace Hamiltonian.
The minisuperspace study of the Liouville theory is recalled in \cref{app:liouville}.
\Cref{app:adm} derives the Hamiltonian of the full Mabuchi action using an ADM parametrization.

\section{Quantum two-dimensional gravity}
\label{sec:quantum}

In this section we recall some properties of the two-dimensional quantum gravity.
It is well-known that the Liouville action corresponds to the effective action for gravity in presence of conformal matter, but when non-conformal matter is present other functionals contribute, such as the Mabuchi action~\cite{Ferrari:2011:RandomGeometryQuantum, Ferrari:2012:GravitationalActionsTwo, Bilal:2014:2DQuantumGravity} (see also~\cite{Ferrari:2013:RandomKahlerMetrics}).
Since the functional integrals for gravity are not well-defined in Minkowski signature we will use the Euclidean signature in this section (under a Wick rotation the Lagrangians are related by $\mc L = - \mc L_E$).

\subsection{Classical actions and partition functions}
\label{sec:quantum:definitions}

Let $\mc M$ be a $2$-dimensional space with metric $g_{\mu\nu}$ with Euclidean signature and whose coordinates are denoted by $\sigma^\mu$.
The matter fields are collectively denoted by $\psi$.

The classical theory is described by the action
\begin{equation}
	\label{quant:action:total}
	S[g, \psi] = S_\mu[g] + S_m[g, \psi]
\end{equation} 
where $S_\mu$ is the cosmological constant action, proportional to the area $A[g]$ of the surface
\begin{equation}
	\label{quant:action:cosmo}
	S_\mu[g] = \mu \int \dd^2 \sigma\, \sqrt{\abs{g}}
		= \mu\, A[g]
\end{equation} 
The Einstein--Hilbert term is not included because it is a topological invariant
\begin{equation}
	S_{\text{EH}}[g] = \int \dd^2 \sigma\, \sqrt{g}\, R
		= 4\pi \chi, \qquad
	\chi = 2 - 2 h
\end{equation} 
with $\chi$ being the Euler number and $h$ the genus of the surface.
The energy--momentum tensor associated to $S_m$ is defined by
\begin{equation}
	\label{quant:eq:energy-tensor}
	T^{(m)}_{\mu\nu} = - \frac{4\pi}{\sqrt{g}} \frac{\var S_m}{\var g^{\mu\nu}}.
\end{equation} 
The classical action is in general not invariant under Weyl transformations
\begin{equation}
	\label{quant:eq:transf-weyl}
	g_{\mu\nu} = \e^{2\omega} g'_{\mu\nu}.
\end{equation} 
If it is invariant then its energy--momentum tensor \eqref{quant:eq:energy-tensor} is traceless $T^{(m)} = 0$.

The total partition function $Z$ and the matter partition function $Z_m$ are
\begin{subequations}
\begin{gather}
	\label{quant:eq:Z-full}
	Z = \frac{1}{\Omega} \int \dd_g g_{\mu\nu}\, \e^{- S_\mu[g]}\, Z_m[g], \\
	\label{quant:eq:Z-matter}
	Z_m[g] = \e^{- S_{\text{eff}}[g]}
		= \int \dd_g \psi\, \e^{- S_m[g, \psi]}.
\end{gather}
\end{subequations}
where $\Omega$ is the volume of the diffeomorphism group and $S_{\text{eff}}$ is the metric effective action induced from the matter.
The subscript $g$ of the measures indicates that they depend on the metric $g$.
The quantum energy--momentum tensor is derived from the effective action
\begin{equation}
	\label{quant:eq:quant-energy-tensor}
	\mean{T^m_{\mu\nu}} = - \frac{4\pi}{\sqrt{\abs{g}}}\, \frac{\var S_{\text{eff}}}{\var g^{\mu\nu}}.
\end{equation}

\subsection{Conformal gauge}
\label{sec:quantum:conformal}

At the quantum level gravity becomes dynamical in two dimensions due to the fact that the quantum fluctuations of the matter fields induce an effective action $S_{\text{eff}}$ for the metric; unfortunately the computation cannot be performed in general.
In order to make progress one can adopt a gauge to fix the diffeomorphisms and the simplest choice is the conformal gauge where the metric is decomposed into a dynamical conformal factor $\phi$ -- the Liouville field -- and a background metric $g_0$
\begin{equation}
	\label{quant:eq:conformal-gauge}
	g = \e^{2\phi} g_0.
\end{equation} 
This gauge fixing cancels the factor $\Omega$ and leads to a Jacobian $Z_{\text{gh}}[g]$ represented by Faddeev--Popov ghosts
\begin{equation}
	\Omega^{-1} \, \dd_g g_{\mu\nu} = \dd\tau \, \dd_g \phi\, Z_{\text{gh}}[g]
\end{equation} 
The moduli $\tau$ are complex continuous parameters that classify the Riemann surfaces of a given genus that are not conformally equivalent (for example, $\tau$ is in the fundamental domain of the upper-half plane when $\mc M = T^2$ is the torus).
They will play no role in the rest of the discussion and as such they will be ignored.
Similarly the ghost contribution will be lumped inside the matter partition function.
In order to distinguish quantities computed in the metric $g$ and $g_0$, the ones associated with the latter will have an index $0$: for example $A_0$ is the area for the metric $g_0$ while $A$ is the area for the metric $g$.

The next step is to express the full partition function \eqref{quant:eq:Z-full} in terms of the matter partition function \eqref{quant:eq:Z-matter} in the background metric
\begin{equation}
	Z = Z_m[g_0] \int \dd_g \phi \, \e^{- S_\mu[g_0, \phi]}\, \e^{- S_{\text{grav}}[g_0, \phi]}
\end{equation}
where the gravitational Wess--Zumino effective action for the Liouville field has been defined by
\begin{equation}
	\label{quant:action:quantum-phi}
	S_{\text{grav}}[g_0, \phi] \equiv S_{\text{grav}}[g, g_0]
		= S_{\text{eff}}[g] - S_{\text{eff}}[g_0]
		= - \ln \frac{Z_m[g]}{Z_m[g_0]}.
\end{equation} 
The main interest of the conformal gauge is that the dynamics of the $\phi$ and $\psi$ fields are totally decoupled (as long as one ignores the moduli), as is obvious in the above expression, and both sectors are field theories on a fixed curved space.
Correlation functions are then simply the product of the correlation functions in each sector and the problem simplifies drastically.

There is a freedom in the decomposition \eqref{quant:eq:conformal-gauge} of the physical metric into a conformal factor and a background metric: this amounts to the existence of an emergent Weyl symmetry
\begin{equation}
	\label{quant:sym:emerging-weyl}
	g_0 = \e^{2\omega} g'_0, \qquad
	\phi = \phi' - \omega.
\end{equation} 
The latter is equivalent to the diffeomorphisms in terms of the physical metric and thus should be preserved.
The most important consequence is that the total action (Liouville and matter fields) should be a CFT on the background $g_0$.

Since every orientable $2$-dimensional manifold is Kähler, another parametrization of the metric $g$ is possible in terms of the Kähler potential $K$.
For this reason it is possible to trade the Liouville mode for the Kähler potential $K$~\cite{Ferrari:2011:RandomGeometryQuantum, Ferrari:2012:GravitationalActionsTwo, Bilal:2014:2DQuantumGravity}
\begin{equation}
	\label{quant:eq:relation-phi-K}
	\e^{2\phi} = \frac{A}{A_0} \left(1 + \frac{A_0}{2}\, \lap_0 K \right),
\end{equation} 
$\lap_0$ being the Laplacian associated to the metric $g_0$.\footnotemark{}
\footnotetext{%
	The different sign compared with~\cite{Ferrari:2011:RandomGeometryQuantum, Ferrari:2012:GravitationalActionsTwo, Bilal:2014:2DQuantumGravity} is that they denote by $\lap$ the positive Laplacian which corresponds to $- \lap$ in our conventions.
	Another difference is that we will normalize the functionals by $4\pi$.
}%
For a given pair $(A, K)$ this relation defines the $\phi$ uniquely (up to constant shift of $K$), and positivity of the exponential implies the inequality
\begin{equation}
	\label{quant:eq:inequality-lap-K}
	\lap_0 K > - \frac{2}{A_0}.
\end{equation}
The Kähler parametrization is very convenient because it can be used to write local actions that would otherwise be non-local in terms of the Liouville field (in the same way that actions non-local in terms of the curvature can be made local in terms of the Liouville field).
The main drawback of this formalism is that it forces to work at fixed area, and subtleties may originate from this as we will find in the next section.

\subsection{Effective actions}

Various functionals appear in the gravitational action \eqref{quant:action:quantum-phi}, the most notable ones (beside the area functional) are:\footnotemark{} the Liouville functional, the Mabuchi functional and the Aubin--Yau functional.
\footnotetext{%
	It is expected that, in general, other functionals are present.
}%
The first one is well-known and describes the effective action when gravity is coupled to conformal matter only, while the other two appear when it is coupled to massive matter~\cite{Ferrari:2011:RandomGeometryQuantum, Ferrari:2012:GravitationalActionsTwo}.
Note that recently all these functionals have been used in the description of the fractional quantum Hall effect~\cite{Ferrari:2014:FQHECurvedBackgrounds, Can:2015:GeometryQuantumHall}.
The next subsections will describe and compare the properties of the Liouville and Mabuchi functionals in order to infer possible properties of the minisuperspace approximation.

\subsubsection{Cosmological constant action}

In the conformal gauge \eqref{quant:eq:conformal-gauge}, the cosmological constant action \eqref{quant:action:cosmo} (also called the area functional) reads
\begin{equation}
	\label{quant:action:cosmo-conf}
	S_\mu = \mu \int \dd^2 \sigma\, \sqrt{g_0}\, \e^{2\phi}.
\end{equation} 
The cosmological constant $\mu$ can receive quantum corrections and its value may differ from the classical one, but we keep the same symbol.
The associated energy--momentum tensor is
\begin{equation}
	\label{quant:eq:cosmo-T}
	T^{(\mu)}_{\mu\nu} = 2 \pi \mu \, g_{0\mu\nu} \, \e^{2\phi}, \qquad
	T^{(\mu)} = 4 \pi \mu \, \e^{2\phi},
\end{equation} 
while the variation of the action is
\begin{equation}
	\label{quant:eq:cosmo-var-S}
	\frac{1}{\sqrt{g_0}} \frac{\var S_\mu}{\var \phi} = 2 \mu \, \e^{2\phi}.
\end{equation}

\subsubsection{Liouville action}
\label{sec:quantum:actions:liouville}

The Liouville action is~\cite{Polyakov:1981:QuantumGeometryBosonic}
\begin{equation}
	\label{quant:action:liouville}
	S_L = \frac{1}{4\pi} \int \dd^2\sigma\; \sqrt{g_0}\, \Big( g_0^{\mu\nu} \pd_\mu \phi \, \pd_\nu \phi + R_0 \, \phi \Big).
\end{equation} 
In the gravitational action it enters with a coefficient $1/b^2$ proportional to the central charges of the matter plus ghosts, and it is the only contribution besides the cosmological constant if the matter is a CFT.
The variation of the action is
\begin{equation}
	\label{quant:eq:liouville-var-S}
	\frac{4\pi}{\sqrt{g_0}} \frac{\var S_L}{\var \phi} = R_0 - 2 \lap_0 \phi
		= \e^{2\phi} R.
\end{equation} 
The trace of the energy--momentum tensor reads
\begin{equation}
	\label{quant:eq:liouville:T-trace-eom}
	T^{(L)} = - \lap_0 \phi,
\end{equation} 
and the latter shows that the Liouville theory is a CFT since one can add an improvement term to the action to set the trace to zero~\cite{Polchinski:1988:ScaleConformalInvariance}.
This is in agreement with the fact that the matter is conformal, which implies that the Liouville action itself should also be conformal since the combined theory should be conformally invariant.

Considering the Liouville theory defined at fixed area in the case where the matter is a CFT, the only contribution to the gravitational action is $\frac{1}{b^2}\, S_L$ (since the cosmological constant $S_\mu = \mu A$ is fixed) and the equation of motion reads~\cite[sec.~2]{Zamolodchikov:1996:StructureConstantsConformal}
\begin{equation}
	\label{quant:eom:liouville-A}
	\frac{4\pi}{\sqrt{g_0}} \frac{\var S_L}{\var_A \phi} = 0
	\quad \Longrightarrow \quad
	\lap R = 0
	\quad \Longrightarrow \quad
	R = \frac{4\pi\chi}{A}
\end{equation} 
where the subscript $A$ on the variation indicates that only variations of $\phi$ which keep the area fixed are considered.
At variable area the cosmological constant enters into the expression of the gravitational action and the equation of motion is found to be
\begin{equation}
	\label{quant:eom:liouville-mu}
	R_0 - 2 \lap_0 \phi = - 8 \pi \mu b^2 \e^{2 \phi}
	\quad \Longrightarrow \quad
	R = - 8 \pi \mu b^2
\end{equation} 
by combining \eqref{quant:eq:cosmo-var-S} and \eqref{quant:eq:liouville-var-S}.
Identifying \eqref{quant:eom:liouville-A} and \eqref{quant:eom:liouville-mu} leads to the relation
\begin{equation}
	\label{quant:eq:relation-mu-A}
	- 8 \pi \mu b^2 = \frac{4\pi\chi}{A}.
\end{equation} 
This relation also results from integrating \eqref{quant:eom:liouville-mu} over the manifold and it can be seen in correlation functions upon performing the Laplace transform (which means that it holds not only on-shell, see~\cite[sec.~2]{Zamolodchikov:1996:StructureConstantsConformal} for the case $\chi = 2$).
In some way this relation encodes how to pass from the fixed to the variable area expressions in the Liouville case, and one may hope that it generalizes to the case of the Mabuchi action.

It is tempting to make the following identification (at least as a rough analogy)
\begin{equation}
	2 \mu = \frac{\sign \chi}{A}, \qquad
	b^2 \sim \abs{\chi}.
\end{equation} 
This suggests that $b$ and $\chi$ may play analogous roles\footnotemark{}; we will come back on this point later.
\footnotetext{%
	This identification can also be motivated by comparing the minisuperspace results of the Liouville and Mabuchi actions.
}%

\subsubsection{Mabuchi action}
\label{sec:quantum:actions:mabuchi}

In the Kähler parametrization the Mabuchi action\footnotemark{} reads~\cite{Ferrari:2011:RandomGeometryQuantum, Ferrari:2012:GravitationalActionsTwo, Bilal:2014:2DQuantumGravity, Can:2015:GeometryQuantumHall}
\footnotetext{%
	We normalize the action by $4\pi$ with respect to~\cite{Ferrari:2011:RandomGeometryQuantum, Ferrari:2012:GravitationalActionsTwo, Bilal:2014:2DQuantumGravity}.
	Note that the Kähler potential of~\cite{Can:2015:GeometryQuantumHall} corresponds to the one of the previous references divided by $A$.
}
\begin{equation}
	\label{quant:action:mabuchi}
		S_M = \frac{1}{4\pi} \int \dd^2 \sigma\, \sqrt{g_0}\; \left(
				- \pi \chi\, g_0^{\mu\nu} \pd_\mu K \, \pd_\nu K
				+ \left( \frac{4\pi \chi}{A_0} - R_0 \right) K
				+ \frac{4}{A}\, \phi \, \e^{2\phi}
			\right)
\end{equation}
where the last term can be expressed in terms of $K$ through \eqref{quant:eq:relation-phi-K}.
It was shown in~\cite{Ferrari:2012:GravitationalActionsTwo, Ferrari:2011:RandomGeometryQuantum} that the Mabuchi action appears in the gravitational action of a massive scalar field (at leading order in a small mass expansion).
The properties of this action have been further studied in~\cite{Bilal:2014:2DQuantumGravity} (see also~\cites{Ferrari:2014:FQHECurvedBackgrounds}[app.~F]{Can:2015:GeometryQuantumHall}).

It is not known whether the Mabuchi action defines a CFT but it seems unlikely to be the case: the non-conformal matter action is not invariant by itself while the total action should be invariant, and hence the non-invariance of the matter action should be compensated by the transformation of the Mabuchi action.

The equation of motion for $K$ (or for $\phi$ at fixed $A$) is
\begin{equation}
	\label{quant:eom:mabuchi}
	R = \frac{4\pi \chi}{A}.
\end{equation} 
It is the same equation as the one of Liouville \eqref{quant:eom:liouville-A}.

\subsection{Rescaling the Mabuchi action}
\label{sec:quantum:rescaling}

In order to prepare the study of the minisuperspace it is necessary to rescale the Kähler potential and the Mabuchi action
\begin{equation}
	\label{quant:eq:rescaling-mabuchi}
	K = \frac{\tilde K}{\pi\chi}, \qquad
	S_M = \frac{\tilde S_M}{\pi\chi}
\end{equation} 
such that the action reads
\begin{equation}
	\label{quant:action:mabuchi-rescaled}
		\tilde S_M = \frac{1}{4\pi} \int \dd^2 \sigma\, \sqrt{g_0}\; \left(
				- g_0^{\mu\nu} \pd_\mu \tilde K \, \pd_\nu \tilde K
				+ \left( \frac{4\pi \chi}{A_0} - R_0 \right) \tilde K \\
				+ \frac{4\pi \chi}{A}\, \phi \, \e^{2\phi}
			\right)
\end{equation}
and the relation \eqref{quant:eq:relation-phi-K} becomes
\begin{equation}
	\label{quant:eq:relation-phi-K-rescaled}
	\e^{2\phi} = \frac{A}{A_0} \left(1 + \frac{A_0}{2\pi \chi}\, \lap_0 \tilde K \right).
\end{equation} 
In the rest of the paper we will omit the tildes on $K$ and $S_M$.
Note that, introducing the dependence in the area, the above action can also be written as
\begin{equation}
	\label{quant:action:mabuchi-decomposition}
	S_M[A] = S_M[A_0] + \frac{\chi}{2} \ln \frac{A}{A_0}.
\end{equation} 

This rescaling requires explanations since it is singular for $\chi = 0$ (genus $1$ surfaces), which is precisely the case we will be looking at in the following sections.
We want to argue that this rescaling is necessary in order to get a consistent result:
\begin{enumerate}
	\item The first point is that the kinetic term of \eqref{quant:action:mabuchi} vanishes for $\chi = 0$ which indicates possible pathologies.
	On the other hand the action \eqref{quant:action:mabuchi-rescaled} is canonically normalized.\footnotemark{}%
	\footnotetext{%
		Up to a minus sign that we expect to be also an artifact of the Kähler parametrization.
	}%
	
	\item Despite the fact that the relation between the fields is singular for $\chi \to 0$, the equations of motion, the Hamiltonian and the spectrum are well-defined even in the limit $\chi \to 0$.
	
	\item There are various instances where the action and/or the fields are rescaled by a parameter that tends to zero.
	This procedure is used to extract meaningful information when the details of the system are smeared in the limit we are taking such that one needs to "zoom".
	Another way to phrase this effect is that most fluctuations of the fields disappear in the corresponding limit, and only the ones scaling appropriately with the parameter remain, but they are visible only after rescaling.
	Some well-known examples are (in most of them the parameter ):
	\begin{itemize}
		\item The most obvious example is related to the semi-classical limit of the path integral which contains a factor $\hbar^{-1}$ in front of the action.
		In order to study the saddle-point approximation it is better to keep this factor like it is, but for other applications it is more useful to rescale the field and coupling constants~\cite{Brodsky:2011:HbarExpansionQuantum}.
		
		\item A similar case is the Yang--Mills gauge theories: the Lagrangian is naturally defined as $L = g^{-2} \tr F^2$~\cite{Weinberg:2005:QuantumTheoryFields-2} but one needs to rescale the gauge field before studying the perturbative expansion in $g$.
		
		\item One can also consider systems where the number of degrees of freedom is taken to be infinite -- for example in large $N$ vector, matrix or tensor models --.
		Without rescaling appropriately the coupling constants and the fields by a (power of) $N$ the dynamics becomes trivial (see for example~\cite{ZinnJustin:1991:ONVectorField, DiFrancesco:1993:2DGravityRandom, Gurau:2011:DoubleScalingLimit, Ferrari:2017:LargeLimitPlanar}).
		Note that in this case $N$ is an integer like $\chi$.
		
		\item A closer example to our problem is the Liouville theory.\footnotemark{}
		\footnotetext{%
			The analogy is not perfect because the parameter in the Liouville theory is continuous while in the Mabuchi case it is discrete.
			But as shown in the previous example there are theories in which the limit is taken for a discrete parameter.
		}%
		Usually the Liouville action with a cosmological constant is written as (in particular when it is studied by itself)
		\begin{equation}
			S'_L = \frac{1}{4\pi} \int \dd^2 \sigma \sqrt{g_0} \left( (\pd\phi)^2 + Q R_0 \phi + 4\pi\mu \, \e^{2b \phi} \right)
		\end{equation} 
		where $Q = 1/b$ or $Q = 1/b + b$ depending on whether one is working with the action of \cref{sec:quantum:actions:liouville} or with the DDK/bootstrap action~\cite{Distler:1989:ConformalFieldTheory, David:1988:ConformalFieldTheories, Zamolodchikov:1996:StructureConstantsConformal}.
		The above action is not well-defined in the semi-classical limit $b \to 0$: for this reason one needs to perform the rescaling
		\begin{equation}
			\phi = \frac{\phi_c}{b}, \qquad
			\mu = \frac{\mu_c}{b^2}, \qquad
			S_L = b^2 S_c.
		\end{equation} 
		This should be compared with \eqref{quant:eq:rescaling-mabuchi}.
		Moreover it should be noted that the semi-classical limit is part of the minisuperspace approximation~\cite[sec.~5]{Zamolodchikov:1996:StructureConstantsConformal} (remember also the comment at the end of \cref{sec:quantum:actions:liouville}).
		
		\item A last simple case is dimensional reduction, where the volume of the additional dimensions are taken to zero: since it multiplies the full action it is necessary to rescale the latter to obtain a non-trivial result.
	\end{itemize}
	
	\item The Kähler formalism itself presents other oddities.
	For example in~\cite{Ferrari:2012:GravitationalActionsTwo} it was found that in the gravitational action the factors multiplying the Mabuchi (and Aubin--Yau) actions depends on $A$.
	As a consequence the equation of motion for the area (necessary to recover the full dynamics with respect to the Liouville field) contains the actions themselves, which is odd.
	More generally it is strange that what would be coupling constants in standard cases depend on a parameter that is integrated over in the functional integral.
	Another difficulty is to compute the energy--momentum tensor: taking the Liouville mode and the background metric as the independent variables, the relation \eqref{quant:eq:relation-phi-K} implies that the variation of $K$ in terms of $g_0$ does not vanish (and similarly for $A$ and $A_0$) and thus the variation of the action in terms of $g_0$ is involved.
	
	\item Adding the cosmological constant term $\mu A$ and using the expression \eqref{quant:action:mabuchi-decomposition} one directly finds the equation of motion for the area to be
	\begin{equation}
		\mu + \frac{\chi}{2 A} = 0,
	\end{equation} 
	which corresponds to \eqref{quant:eq:relation-mu-A}.
	Then by plugging this result into \eqref{quant:eom:mabuchi} one finds the same equation than \eqref{quant:eom:liouville-mu}, in the same way that \eqref{quant:eom:mabuchi} was matching \eqref{quant:eom:liouville-mu}.
	Due to the comments below \eqref{quant:eq:relation-mu-A} it is possible that this relation holds at the level of the functional integral.
	This will be used in the next section to infer the possible minisuperspace action, where we will find other support for this procedure.
	
	\item The action \eqref{quant:action:mabuchi-rescaled} contains only the geometric quantities
	\begin{equation}
		\bar R = \frac{4\pi\chi}{A}, \qquad
		\bar R_0 = \frac{4\pi\chi}{A_0},
	\end{equation} 
	and this is also true of the factors in front of the Mabuchi action in~\cite{Ferrari:2012:GravitationalActionsTwo} (in agreement with the comment at the bottom of p.~21 of~\cite{Ferrari:2012:GravitationalActionsTwo}).
	According to the previous point this would mean that every instance of $\chi / A$ could be replaced by $- \mu$ and this would remove the ambiguities described above (with this interpretation the apparent divergences discussed below (4.21) of~\cite{Ferrari:2012:GravitationalActionsTwo} would be an artifact of the formulation).
	
	\item The form of the minisuperspace approximation of the unscaled action \eqref{quant:action:mabuchi} can be found in \cref{app:adm} and is seen to not give a meaningful result.
	
	\item One could have considered to rescale by $- \pi\chi$ instead in order to make the kinetic term positive definite, but one would find that the Hamiltonian is not positive definite -- see \eqref{mab:eq:hamiltonian-mabuchi-adm} and the comment below \eqref{mini:eom:mabuchi-mu} -- and the identification \eqref{quant:eq:relation-mu-A} would not hold for the potential.
	Moreover as argued in point 5) one should not take too seriously the negative sign in front of the kinetic term since the coupling constant is also negative for $\chi < 0$: then the replacement \eqref{quant:eq:relation-mu-A} would make the combination positive for all genus.
\end{enumerate}
Even if none of these arguments is sufficiently rigorous to prove alone that the rescaling is well-defined, the convergence of these arguments gives support to this idea and points more toward the fact that the various pathologies are not genuine but rather due to the formalism.
Since there is no other formalism at our disposition we will use the action \eqref{quant:action:mabuchi-rescaled} as our starting point.

Finally the action \eqref{quant:action:mabuchi-rescaled} will be modified a last time to
\begin{equation}
	\label{quant:action:mabuchi-rescaled-bdy}
		S_M = \frac{1}{4\pi} \int \dd^2 \sigma\, \sqrt{g_0}\; \left[
				- g_0^{\mu\nu} \pd_\mu K \, \pd_\nu K
				+ \left( \frac{4\pi \chi}{A_0} - R_0 \right) K \\
				+ \frac{2\pi \chi}{A}\, (2 \phi - 1) \, \e^{2\phi}
			\right]
\end{equation}
where a trivial term has been added.
In terms of the Liouville mode it means that one can shift the field $\phi$ by a constant term without changing anything, while in terms of the Kähler potential it becomes a constant term and a boundary term.
Moreover the addition of this term makes the variation of the action better defined since it cancels a boundary term proportional to the normal derivative of $\delta K$, which does not vanish (in the same way that one is adding a Gibbons--Hawking--York term in general relativity).
The Hamiltonian of this action is computed in \cref{app:adm}.

\section{Computations of the minisuperspace Hamiltonian}
\label{sec:minisuperspace}

The goal of this section is to motivate the action we proposed in~\cite{deLacroix:2016:MabuchiSpectrumMinisuperspace} for describing the minisuperspace approximation of the Mabuchi action \eqref{quant:action:mabuchi-rescaled-bdy} in order to proceed to its canonical quantization in the next section.
This action\footnotemark{} reads (in Lorentzian signature)
\footnotetext{%
	Note that for $\varpi = \frac{\ddot K}{8\pi \mu}$ the second term of this action corresponds to the one of the flat (or BMS) Liouville theory in the minisuperspace approximation.
	The latter corresponds to the asymptotic theory of $3d$ Minkowski $M_3$ in the same sense that the usual Liouville theory is the asymptotic theory of $\group{adS}_3$~\cite{Barnich:2012:FlatLimitThree, Barnich:2014:2DFieldTheory}.

}%
\begin{equation}
	\label{mini:action:mabuchi}
	S_M = - \frac{1}{2} \int \dd t \left[
			\dot K^2
			- \ddot K \ln \left( \frac{\ddot K}{4\pi \mu} \right)
			+ \ddot K
		\right]
\end{equation}
together with the relation between $\phi$ and $K$
\begin{equation}
	\label{mini:eq:relation-phi-K}
	\e^{2\phi} = \frac{\ddot K}{4\pi\mu},
\end{equation}
and its Hamiltonian is equal to the one of the Liouville theory (in the minisuperspace approximation)
\begin{equation}
	\label{mini:eq:hamiltonian-mabuchi}
	H_M = \frac{\Pi^2}{2} + 2 \pi\mu \, \e^{2\phi},
\end{equation} 
$\Pi$ being the conjugate momentum of $\phi$.
The equation of motion for $K$ derived from \eqref{mini:action:mabuchi} reads
\begin{equation}
	\label{mini:eom:mabuchi-mu}
	\ddot \phi = - 4\pi \mu \, \e^{2\phi}
\end{equation} 
after replacing $K$ by $\phi$ with \eqref{mini:eq:relation-phi-K}.
This is the minisuperspace Liouville equation of motion resulting from \eqref{quant:eom:liouville-mu} and it corresponds to the expected variable area minisuperspace approximation of \eqref{quant:eom:mabuchi} (following the comments in \cref{sec:quantum:rescaling}).
The Hamiltonian equation of motion for $\phi$ derived from \eqref{mini:eq:hamiltonian-mabuchi} clearly reproduces this equation.

Due to various pathologies of the formalism (explained in \cref{sec:quantum:rescaling,sec:minisuperspace:approx}) we have not been able to give a rigorous proof that \eqref{mini:action:mabuchi} is the correct minisuperspace Lorentzian action.
Nonetheless we present three computations of the Hamiltonian \eqref{mini:eq:hamiltonian-mabuchi} (from which the Lagrangian \eqref{mini:action:mabuchi} can be derived through a Legendre transformation) that all rely on different (mild) assumptions and for this reason we believe that together they provide a support for our conjecture.
Moreover since the action \eqref{mini:action:mabuchi} reproduces the main characteristics of the Mabuchi action \eqref{quant:action:mabuchi} (standard kinetic term for $K$ and potential in $\phi \, \e^{2\phi}$) it is expected to capture the main features of the zero-mode dynamics in the Mabuchi theory.
Hence, even if a rigorous derivation can be performed only by starting with a variable area action which is not known, one is still able to make progress.

The first subsection explains the various subtleties of the minisuperspace approximation while the other ones present the different derivations.

\subsection{Minisuperspace approximation}
\label{sec:minisuperspace:approx}

The minisuperspace approximation consists in studying only time-dependent Kähler potential and Liouville mode 
\begin{equation}
	\phi(t, \sigma) = \phi(t), \qquad
	K(t, \sigma) = K(t).
\end{equation} 
In order to single out a globally defined time direction (global hyperbolicity) for the Hamiltonian formalism, the background spacetime is taken to be a cylinder $I \times S^1$ where $I$ is an interval of length $T$.
This cylinder is obtained from the torus $T^2$ by unwrapping one of its dimension.
This direct product structure implies that the spacetime is flat
\begin{equation}
	\label{mini:eq:flat-limit}
	\chi = 0, \qquad
	g_0 = \eta, \qquad
	t \in \left[ - \frac{T}{2}, \frac{T}{2} \right], \qquad
	\sigma \in [0, 2\pi).
\end{equation} 
The physically-relevant case is when time is non-compact with $T \to \infty$ and $I = \R$ (the infinite cylinder can also be obtained by taking the radius of one of the torus circle to infinity).
Unfortunately several difficulties arise from the fact that the Mabuchi action is formulated at fixed area, and that it depends on $\chi$ and $A_0 = 2\pi \, T$.
For this reason one needs to be careful when taking the limits.

The dynamical variables in the fixed area formalism are $K$ and $A$ and they do not include $A_0$.
Since the background metric $g_0$ results from a gauge choice, it can be chosen such that $A_0$ has some specific value (in particular the area $A_0$ does not appear in the equation of motion).
For these different reasons it is expected that one can take $A_0 \to \infty$ (corresponding to $T \to \infty$), independently of the value of $A$ which can be kept finite.
Note that this limit is taken by changing the ranges of the coordinates as described above and not the components of the metric: for this reason $g_0$ is kept fixed while taking the limit.\footnotemark{}
\footnotetext{%
	This point is similar to the question of whether the moduli describing a Riemann surfaces appear in the definition of the coordinate ranges or in the metric~\cite{Polchinski:2005:StringTheory-1}.
}%
This may introduce some spurious singularity in $\phi$ but this will be of no importance for our study.
In any case working at variable area is necessary in order to have a non-trivial dynamics in the minisuperspace: starting with the Liouville action \eqref{quant:action:liouville} one finds that the minisuperspace Hamiltonian reduces to a free field Hamiltonian.
Below we will find the same result for the Mabuchi minisuperspace Hamiltonian.

In a second step we impose that the curvature vanishes since the spacetime is flat, $R_0 = 0$.
A potential problem can arise because $\chi = 0$ for flat space, but the Lagrangian is singular in this case.
Two different solutions are possible.
The first one is to consider a singular $\phi$ such that
\begin{equation}
	4\pi\chi = \int \dd^2 \sigma\, \sqrt{\abs{g}}\, R
		\neq \int \dd^2 \sigma\, \sqrt{\abs{g_0}}\, R_0,
\end{equation} 
in which case $R_0 = 0$ does not imply $\chi = 0$.
One is then forced to work with patches in order to deal with the singularity of $\phi$: as explained above we do not work directly with the value of $\phi$ in this paper and this should be of no consequence.
The second solution is simpler: from the relation
\begin{equation}
	\label{mini:eq:relation-R0-A0}
	R_0 A_0 = 4\pi \chi,
\end{equation} 
which is valid for constant $R_0$, one sees that $\chi$ can take a non-zero value if one takes the limits $A_0 \to \infty$ and $R_0 \to 0$ simultaneously such that the product is constant (this is a form of double scaling limit).
We will adopt this view in \cref{sec:minisuperspace:deriv-ostrogradski} and we will formally work with $\chi \neq 0$.

This can be rephrased in terms of the torus moduli $\tau = \tau_1 + i \tau_2$ (living in the fundamental domain of the upper-half plane)~\cite[sec.~7.1]{Polchinski:2005:StringTheory-1}.
The flat metric in \eqref{mini:eq:flat-limit} can be written as
\begin{equation}
	\dd s^2
		= \dd t^2 + \dd x^2
\end{equation} 
where the coordinates are periodically identified
\begin{equation}
	(t, x) \sim (t, x + 2\pi)
		\sim (t + 2\pi \tau_2, x + 2\pi \tau_1).
\end{equation} 
In this case the torus is described as a cylinder of length $T = 2\pi \tau_2$ whose ends are identified with a twist of $2\pi \tau_1$.
The decompactification of the torus to the cylinder corresponds to the limit $\tau_2 \to \infty$.
Hence the minisuperspace limit corresponds to a specific corner of the moduli space.

Another motivation for keeping $\chi$ arbitrary until the end is the fact that the operations of taking a limit in the Lagrangian or in some quantity computed from it may not commute.
The Liouville theory again provides an example: it is well-known that one should not set $R_0 = 0$ in the Lagrangian before computing the energy--momentum tensor for the flat space case $g_0 = \eta$ since the variation of this term gives a non-vanishing contribution in the limit $R_0 \to 0$.
So one should avoid to take limits directly in the Lagrangian if one is not sure of the effect this will have when computing other quantities.

Finally the question of the Wick rotation needs to be addressed since a positive definite Hamiltonian requires the signature to be Lorentzian.
Note that compact spacetimes with Lorentzian signature are perfectly well-defined and it can be convenient to use them at intermediate stages of computations.
For example it is frequent in QFT to consider “spacetime in a box” in order to regulate IR divergences, before taking the infinite limit volume.
Moreover the equation of motion for the Liouville mode has the same form \eqref{quant:eom:liouville-mu} in both Euclidean and Lorentzian signatures.
Hence the relation \eqref{quant:eq:relation-mu-A} also holds and indicates that classical solutions have finite area $A$ even in Lorentzian signature except possibly if $\chi = 0$ at the same time.
For these reasons it is fine to first perform the Wick rotation of the action and later to consider the infinite area limit.

\subsection{First derivation: infinite area and flat limits}
\label{sec:minisuperspace:deriv-limits}

Gathering all the previous elements, the relation \eqref{quant:eq:relation-phi-K-rescaled} between the Liouville and Kähler fields becomes
\begin{equation}
	\label{mini:eq:relation-phi-K-A}
	\e^{2\phi} = - \frac{A}{2\pi \chi}\, \ddot K.
\end{equation}
and the minisuperspace action of the Mabuchi action \eqref{quant:action:mabuchi-rescaled-bdy} reads
\begin{equation}
	\label{mini:action:mabuchi-A}
	S_M = - \frac{1}{2} \int \dd t \left[
			\dot K^2
			- \ddot K \ln \left( - \frac{A}{2\pi \chi}\, \ddot K \right)
			+ \ddot K \right].
\end{equation}
The overall minus sign comes from the Lorentzian signature and we have set $R_0 = 0$ while the integration over the spatial direction has provided a factor $2\pi$.
It is straightforward to check that the variation of \eqref{mini:action:mabuchi-A} agrees with the minisuperspace approximation of \eqref{quant:eom:mabuchi}
\begin{equation}
	\label{mini:eom:mabuchi}
	\ddot \phi = \frac{2\pi \chi}{A}\, \e^{2\phi}.
\end{equation} 

Since the action \eqref{mini:action:mabuchi-A} does not depend on $K$ it is possible to reduce it to a first order action by considering $\dot K$ as the canonical variable.\footnotemark{}
\footnotetext{%
	In fact the condition $R_0 = \cst$ is sufficient for this to happen in view of the relation \eqref{mini:eq:relation-R0-A0}.
}%
The conjugate momenta $P$ reads
\begin{equation}
	P = \frac{\var S_M}{\var \ddot K}
		= \frac{1}{2} \ln \left( - \frac{A}{2\pi \chi}\, \ddot K \right).
\end{equation}
It is necessary to invert this expression
\begin{equation}
	\label{mini:eq:relation-dK-P}
	\ddot K = - \frac{2\pi \chi}{A}\, \e^{2 P}.
\end{equation} 
in order to compute the Hamiltonian
\begin{equation}
	\label{mini:eq:hamiltonian-mabuchi-dK}
	H_M = P \ddot K - L_M
		= \frac{\dot K^2}{2} - \frac{\pi \chi}{A}\, \e^{2 P}.
\end{equation} 
Comparing the relations \eqref{mini:eq:relation-dK-P} and \eqref{mini:eq:relation-phi-K-A} shows that $P$ can be identified with $\phi$.
Performing a canonical transformation to exchange position and momentum
\begin{equation}
	P = \phi, \qquad
	\dot K = - \Pi,
\end{equation} 
where $\Pi$ is the canonical momentum associated to $\phi$, provides the Hamiltonian
\begin{equation}
	\label{mini:eq:hamiltonian-mabuchi-A}
	H_M = \frac{\Pi^2}{2} - \frac{\pi \chi}{A}\, \e^{2\phi}.
\end{equation} 
It is straightforward to check that the equations of motion \eqref{mini:eom:mabuchi} follow from this Hamiltonian.
At this point it is possible to set $\chi = 0$ (which is well-defined) and to add the cosmological constant term\footnotemark{} to find \eqref{mini:eq:hamiltonian-mabuchi}
\footnotetext{%
	This is equivalent to insert the cosmological constant term in the path integral and to replace the integration over $(K, A)$ by the one over $\phi$.
	Note that the same effect is achieved using the arguments in section~\ref{sec:quantum:actions:liouville} and the replacement \eqref{quant:eq:relation-mu-A}.
}%
\begin{equation}
	\label{mini:eq:hamiltonian-mabuchi-mu}
	H_M = \frac{\Pi^2}{2} + 2\pi \mu \, \e^{2\phi}.
\end{equation} 
In this form the Hamiltonian is explicitly positive definite and it is nothing else but the Liouville Hamiltonian \eqref{liouv:eq:hamiltonian} in the minisuperspace approximation (with $b = 1$ corresponding to the case where the Liouville mode has not been rescaled).

One may be surprised to start with a Lagrangian \eqref{mini:action:mabuchi} containing a negative-definite kinetic term and to end with a positive-definite Hamiltonian \eqref{mini:eq:hamiltonian-mabuchi-mu}.
This is a consequence of the presence of higher-derivatives: the $\Pi^2$ term comes entirely from the $- L_M$ term which explains why it has the correct sign, compared to the standard computation in the absence of higher-derivatives where the first term contributes typically with an opposite sign and is twice bigger.
In particular one can see here that rescaling with $- \pi\chi$ to get a positive-definite kinetic term in the Lagrangian would have lead to a negative-definite Hamiltonian.

The consistency of these computations can be checked by following the same approach with the Liouville action \eqref{quant:action:liouville} at fixed area: the minisuperspace Hamiltonian of the latter is simply $H_L = p^2 / 2$ which coincides with \eqref{mini:eq:hamiltonian-mabuchi-A} when $\chi = 0$, and the full Hamiltonian \eqref{liouv:eq:hamiltonian} is recovered by adding the cosmological constant.

It is interesting to see that the relation between the Liouville mode and the Kähler potential is built-in in the Hamiltonian formalism since the first appears as the conjugate momentum of the latter.
Thanks to this it is not necessary to impose the relation \eqref{mini:eq:relation-phi-K} (for example using the Dirac formalism) nor the corresponding constraint \eqref{quant:eq:inequality-lap-K}.

\subsection{Second derivation: Legendre transformation}
\label{sec:minisuperspace:deriv-legendre}

It has been observed in~\cite{Klevtsov:2011:2DGravityKahler} that the kinetic and potential terms of the Mabuchi action are respectively Legendre dual to the kinetic term of the Liouville action and to the cosmological constant potential (the kinetic terms including the linear piece).
Despite the fact that this relation did not receive any explanation it gives a simple consistency check of the other derivations: we will apply it in the minisuperspace approximation and show that the resulting action corresponds to \eqref{mini:action:mabuchi}.

Let's define from \eqref{quant:action:liouville} and \eqref{quant:action:mabuchi}
\begin{equation}
	T_L = \frac{\dot \phi^2}{2}, \qquad
	V_L = - 2\pi \mu\, \e^{2\phi}
\end{equation} 
(the $2\pi$ in the second term comes from integrating over $S^1$, since the cosmological constant is not normalized) along with the Legendre transforms of these functions
\begin{equation}
	\label{deriv:eq:Mabuchi-dual}
	T_M = \phi \hat\phi - T_L
		= \phi \hat\phi - \frac{\dot \phi^2}{2}, \qquad
	V_M = \phi \hat\phi - V_L
		= \phi \hat\phi + 2\pi \mu\, \e^{2\phi}.
\end{equation} 
We need to extremize the above functions with respect to $\phi$ and plug back the result.

Let's start with $T_L$:
\begin{equation}
	\frac{\var T_M}{\var\phi} = \hat\phi + \ddot \phi
		= 0.
\end{equation} 
Defining $\hat\phi = - \ddot k$ one obtains the solution, and plugging back gives (under the integral)
\begin{equation}
	T_M = - k \ddot k - \frac{\dot k^2}{2}
		= \frac{\dot k^2}{2}.
\end{equation} 

Let's apply the same procedure to $V_L$:
\begin{equation}
	\frac{\var V_M}{\var\phi} = \hat\phi + 4\pi \mu\, \e^{2\phi}
		= 0.
\end{equation} 
The solution reads
\begin{equation}
	\phi = \frac{1}{2} \ln \left( - \frac{\hat\phi}{4\pi \mu} \right),
\end{equation} 
and defining again $\hat\phi = - \ddot k$ one obtains
\begin{equation}
	V_M = \frac{\hat\phi}{2} \left(
			\ln \left( - \frac{\hat\phi}{4\pi \mu} \right) - 1
			\right)
		= - \frac{\ddot k}{2} \left(\ln \left(\frac{\ddot k}{4\pi \mu} \right) - 1 \right).
\end{equation} 
Note the presence of a boundary term.

The final action that we obtain by gathering both terms is\footnotemark{}
\footnotetext{%
	Note that the Liouville terms were in Lorentzian signature whereas the resulting Mabuchi action is in Euclidean signature.
}%
\begin{equation}
	S_M = \frac{1}{2} \int \dd t \left[
			\dot k^2
			- \ddot k \ln \left( \frac{\ddot k}{4 \pi \mu}\, \right)
			+ \ddot k
		\right].
\end{equation} 
This action is the same as \eqref{mini:action:mabuchi} upon the identification $k = K$ and thus it will yield the Hamiltonian \eqref{mini:eq:hamiltonian-mabuchi-mu}.
Note that it naturally incorporates the boundary term from \eqref{quant:action:mabuchi-rescaled-bdy} and the relation \eqref{quant:eq:relation-mu-A}.

\subsection{Third derivation: Ostrogradski formalism}
\label{sec:minisuperspace:deriv-ostrogradski}

In order to generalize the computation of \cref{sec:minisuperspace:deriv-limits} we will consider the case where $A_0 = 2\pi T$ and $R_0$ are kept finite (generalizing the idea that one should not set terms to zero directly in the Lagrangian).
In this case the Lagrangian is of higher order in the derivatives and one needs to use the Ostrogradski formalism (see~\cite{Woodard:2015:TheoremOstrogradsky} for a recent review).
The more general case starting with the full Lagrangian can be found in \cref{app:adm}.

The Mabuchi action (in Lorentzian signature) in the minisuperspace approximation (no spatial dependence) is
\begin{equation}
	S_M = - \frac{1}{2} \int \dd t \left[ \dot K^2
			+ \left( \frac{4\pi \chi}{A_0} - R_0 \right) K
			+ \frac{2 \pi\chi}{A_0} \left( 1 - \frac{A_0}{2 \pi\chi}\, \ddot K \right) \left(
				\ln \frac{A}{A_0} \left( 1 - \frac{A_0}{2 \pi\chi}\, \ddot K \right)
				- 1 \right)
		\right]
\end{equation} 
with the relation
\begin{equation}
	\label{deriv:eq:relation-phi-k}
	\e^{2\phi} = \frac{A}{A_0} \left(1 - \frac{A_0}{2 \pi\chi}\, \ddot K \right).
\end{equation} 

The canonical variables are taken to be $(K, \mc P)$ and $(\dot K, P)$ where the conjugate momenta are
\begin{subequations}
\begin{align}
	P &= \frac{\pd L}{\pd \ddot K}
		= \frac{1}{2} \ln \frac{A}{A_0} \left( 1 - \frac{A_0}{2 \pi\chi}\, \ddot K \right), \\
	\mc P &= \frac{\pd L}{\pd \dot K} - \frac{\dd P}{\dd t}
		= - \dot K - \frac{1}{2} \frac{\dd P}{\dd t}.
\end{align}
\end{subequations}
In particular we can invert the first relation to find $\ddot K$ in terms of the canonical variable
\begin{equation}
	\ddot K = \frac{2\pi\chi}{A} \left( \frac{A}{A_0} - \e^{2 P} \right).
\end{equation} 
Moreover, comparing this expression with \eqref{deriv:eq:relation-phi-k} one finds $P = \phi$.
The Hamiltonian reads
\begin{equation}
	\label{eq:ostrogradski:hamiltonian-mabuchi}
	H = \mc P \dot K + P \ddot K - L
		= \mc P \dot K
			+ \frac{2 \pi\chi}{A_0}\, P
			+ \frac{\dot K^2}{2}
			+ \frac{1}{2} \left(\frac{4\pi \chi}{A_0} - R_0 \right) K
			- \frac{\pi\chi}{A}\, \e^{2 P}.
\end{equation} 

The canonical transformation
\begin{equation}
	P = \phi, \qquad
	\dot K = - \Pi
\end{equation}
can be performed in order to express the Hamiltonian \eqref{eq:ostrogradski:hamiltonian-mabuchi} in terms of the Liouville field
\begin{equation}
	\label{eq:ostrogradski:hamiltonian-mabuchi-phi}
	H = \frac{\Pi^2}{2}
		- \mc P \Pi
		+ \frac{1}{2} \left(\frac{4\pi \chi}{A_0} - R_0 \right) K
		+ \frac{2 \pi\chi}{A_0}\, \phi
		- \frac{\pi\chi}{A}\, \e^{2 \phi}.
\end{equation} 
It is shown in \cref{app:adm} that this minisuperspace Hamiltonian can be obtained as a limit from the full Hamiltonian computed through an ADM parametrization of the metric.

The Hamiltonian is well-defined for $\chi = 0$ and $R_0 = 0$ and it reduces to
\begin{equation}
	H = \frac{\Pi^2}{2} - \mc P \Pi,
\end{equation} 
noting that in this case it is not necessary to take the limit $A_0 \to \infty$.
After performing the canonical transformation
\begin{equation}
	\tilde \Pi = \Pi - \mc P, \qquad
	\tilde{\mc P} = \mc P, \qquad
	\tilde\phi = \phi, \qquad
	\tilde K = \phi + K,
\end{equation} 
the Hamiltonian reads (omitting the tildes)
\begin{equation}
	H = \frac{\Pi^2}{2} - \frac{\mc P^2}{2}.
\end{equation} 
Adding the cosmological constant gives finally
\begin{equation}
	\label{eq:ostrogradski:hamiltonian-mabuchi-mu}
	H = \frac{\Pi^2}{2} + 2 \pi\mu \, \e^{2\phi} - \frac{\mc P^2}{2}.
\end{equation} 
Hence one recovers Liouville Hamiltonian plus a free (ghost) term.
The interpretation of this additional ghost field with respect to the other methods is not clear but it can be expected that it is just an artifact of the fixed area formalism.

\section{Minisuperspace canonical quantization}
\label{sec:quantization}

The minisuperspace approximation is well-suited to determine the Hilbert space of the theory as the latter can be found by studying the dynamics of the zero-mode only, which simplifies greatly the canonical quantization of the action.
Through several changes of variables we argued that the Mabuchi and Liouville Hamiltonians are equal, implying that the Mabuchi spectrum is identical to the Liouville spectrum.
We review the results presented in our letter~\cite{deLacroix:2016:MabuchiSpectrumMinisuperspace}.

The spectrum of Mabuchi theory is determined through the canonical quantization
\begin{equation}
	\Pi \longrightarrow -i\, \frac{\dd}{\dd\phi}.
\end{equation} 
It coincides with the minisuperspace quantization of the Liouville theory~\cite{DHoker:1982:ClassicalQuantalLiouville, Braaten:1984:NonperturbativeWeakcouplingAnalysis, Seiberg:1990:NotesQuantumLiouville} and we highlight the main features, while we refer the reader to the literature for more details (see~\cite{Schomerus:2003:RollingTachyonsLiouville, McElgin:2008:NotesLiouvilleTheory} for recent accounts). 
The stationary Schrödinger equation reads\footnotemark{}
\footnotetext{%
	If the action had not been rescaled by $\pi\chi$ before, it would be equivalent to rescaling the eigenvalues here.
}%
\begin{equation}
	\label{mini:eq:schrodinger-mabuchi-p}
	H_M \psi_p = 2 p^2\, \psi_p,
\end{equation} 
where the definition of the eigenvalue is conventional, and this provides the differential equation
\begin{equation}
	\label{mini:eq:schrodinger-mabuchi}
	\left(- \frac{1}{2}\, \frac{\dd^2}{\dd \phi^2}
		+ 2 \pi\mu \, \e^{2\phi} - 2 p^2 \right) \psi_p(\phi)
	= 0.
\end{equation}
It corresponds to the modified Bessel equation whose solutions are
\begin{subequations}
\begin{align}
	\psi_p(\phi) &= \frac{2 (\pi\mu)^{- i p}}{\Gamma(- 2i p)}\, K_{2i p}(2 \sqrt{\pi\mu}\, \e^\phi) \\
		&\sim_0 \e^{2i p \phi} + R_0(p) \e^{-2i p \phi}.
\end{align}
\end{subequations}
The second linearly independent solution has been removed because it blows up at $\phi \to \infty$ and the normalization has been chosen such that the incoming plane waves have unit coefficient as $\phi \to -\infty$.
The factor
\begin{equation}
	R_0(p) = \frac{\Gamma(2i p)}{\Gamma(-2i p)}\; (\pi\mu)^{- 2i p}
\end{equation} 
is interpreted as a reflection coefficient in the Liouville theory, but its interpretation in terms of the Mabuchi action is not clear.
Moreover it can be seen that wave functions with $\pm p$ are not independent.
\begin{equation}
	\psi_{-p}(\phi) = R_0(-p)\, \psi_p(\phi).
\end{equation} 

Additional constraints such as normalisability are needed in order to restrict the eigenvalues.
In particular the states of the physical Hilbert space need to be (delta-function) normalizable under the canonical inner product.
It can be seen that this condition is fulfilled only for $p \in \R$
\begin{equation}
	\int_{-\infty}^\infty \dd\phi\; \conj{\psi_p}(\phi) \psi_{p'}(\phi)
		= \pi\, \dirac(p - p').
\end{equation} 
It can also be checked that the states with $p \in \R$ form a complete basis.

Since it is not clear if the Mabuchi theory defines a CFT we do not link these eigenvalues to conformal weights as for the Liouville theory (see appendix~\ref{app:liouville}).
In particular the unitarity condition is not clear and hence we do not comment the status of the states with $p \in i\R$ (recall that some of those play a physical role in $2d$ Liouville gravity).

As a consequence the operators and the associated eigenvalues are identical in the Mabuchi and Liouville theories.
On one hand it is not surprising due to the fact that the classical equations of motion are identical, but on the other hand it is highly non-trivial that the very complicated action \eqref{mini:action:mabuchi} reduces to the Liouville Hamiltonian after performing suitable changes of variables.
Without the identification of the momentum $P$ to the Liouville mode it would have been very difficult to extract the wave functions for $\phi$.

Finally the semi-classical limit of the $3$-point function can be read from the minisuperspace
\begin{subequations}
\label{mini:eq:3-point}
\begin{align}
	C_0(p_1, p_2, p_3) &=
		\int_{-\infty}^\infty \dd\phi\; \psi_{p_1}(\phi) \e^{- 2 i p_2 \phi} \psi_{p_3}(\phi) \\
		&= (\pi\mu)^{- 2 \tilde p}\,
			\Gamma(2 \tilde p)
			\prod_i \frac{\Gamma\big((-1)^i 2 \tilde p_i\big)}{\Gamma(2 p_i)}
\end{align}
\end{subequations}
where we defined
\begin{equation}
	2 \tilde p = \sum_i p_i, \qquad
	\tilde p_i = \tilde p - p_i, \qquad
	i = 1, 2, 3.
\end{equation} 
Of course this result agrees with the Liouville theory in the minisuperspace approximation, but discrepancies will certainly appear beyond the semi-classical limit.

\section{Conclusion and discussion}

In this paper, we have given support for the form of the Mabuchi action in the minisuperspace approximation which had been proposed in~\cite{deLacroix:2016:MabuchiSpectrumMinisuperspace}.
Using this action we could derive the spectrum of the Mabuchi theory and show that it is the same as the one of the Liouville theory.
The knowledge of the spectrum provides a natural set of operators for which to compute the correlation functions, a question which was still open.

The next step is to study the minisuperspace of the coupled Liouville--Mabuchi theory, and more particularly by taking into account the precise coefficients related to the matter content.
To this aim, the gravitational action for a massive scalar field on a cylinder computed in~\cite{Bilal:2017:2DGravitationalMabuchi} can be used.

A major goal is to define more completely the pure Mabuchi theory -- and even more importantly the Liouville--Mabuchi theory --, in particular by performing a more rigorous quantization.
A possible approach to this problem would be to design a formalism that allows extending the definition of the functional to variable area.

As indicated previously, the Mabuchi action is not expected to be conformal in order to compensate for the transformation of the matter action.
This makes our result even more intriguing since the semi-classical approximation of the theory is conformal (due to its equivalence with the Liouville minisuperspace Hamiltonian).
Hence it would be interesting to study which quantum effects break the conformal invariance of the semi-classical Mabuchi theory.
One possibility is that additional states break the conformal invariance but decouple semi-classically.\footnotemark{}
\footnotetext{%
    This may be similar to what happens in the SYK model where the infrared regime develops an emergent conformal symmetry~\cite{Maldacena:2016:CommentsSachdevYeKitaevModel, Polchinski:2016:SpectrumSachdevYeKitaevModel, Dartois:2017:Conformality1NCorrections}.
    In connection, see~\cite{Bojowald:2015:MinisuperspaceModelsInfrared} for a description of the minisuperspace approximation as an infrared cut-off.
}%
Another possibility is that the coupling of the Mabuchi action to the matter will break the conformal symmetry: then the prefactor of the action depends on the area instead of being constant -- and thus different terms may dominate -- and the moduli may also play a role.\footnotemark{}
\footnotetext{%
    We thank an anonymous referee for suggesting these possibilities.
}%
This calls for an exact quantization of the theory.

We hope to come back to these topics in future works.

\section*{Acknowledgements}

We would like to thank Costas Bachas, Frank Ferrari, Semyon Klevtsov, Vincent Lahoche, Raoul Santachiara and Jean-Bernard Zuber for useful discussions.
We are particularly grateful to Adel Bilal and Ashoke Sen for carefully reading the draft of the manuscript.
C.L.\ and H.E.\ would like to thank the Harish--Chandra Research Institute (Allahabad, India) for its hospitality during part of this work, and H.E. acknowledges support from Cefipra under project 5204-4.
The work of E.E.S., made within the \textsc{Labex Ilp} (reference \textsc{Anr–10–Labx–63}), is supported by French state funds managed by the \emph{Agence nationale de la recherche}, as part of the program \emph{Investissements d'avenir} under the reference \textsc{Anr–11–Idex–0004–02}.

\appendix

\section{Liouville theory: minisuperspace}
\label{app:liouville}

In this section we recall the main formulas for the minisuperspace analysis of the Liouville theory \eqref{quant:action:liouville} with a cosmological constant \eqref{quant:action:cosmo-conf} (we will denote the sum by $S_L$ for simplicity)~\cite{DHoker:1982:ClassicalQuantalLiouville, Seiberg:1990:NotesQuantumLiouville, McElgin:2008:NotesLiouvilleTheory}.
In the minisuperspace approximation
\begin{equation}
	\phi = \phi(t), \qquad
	g_0 = \eta,
\end{equation} 
the action reads
\begin{equation}
	S_L = \int \dd t\, \left( \frac{\dot \phi^2}{2} - 2\pi \mu \, \e^{2b\phi} \right).
\end{equation}
The conjugate momentum
\begin{equation}
	p = \frac{\var S_L}{\var \dot\phi}
		= \dot \phi
\end{equation} 
is used to construct the Hamiltonian
\begin{equation}
	\label{liouv:eq:hamiltonian}
	H_L = p \dot \phi - L
		= \frac{p^2}{2} + 2\pi \mu \, \e^{2b\phi}.
\end{equation} 

We do not repeat the analysis of the quantization given in \cref{sec:quantization}.
In order to interpret the spectrum it is necessary to bring the theory back to the plane.
The Hamiltonian on the latter is given by the dilatation operator $L_0 + \bar L_0$ and the associated wave functions are solutions of
\begin{equation}
	(L_0 + \bar L_0) \psi_\Delta = 2 \Delta\, \psi_\Delta
\end{equation} 
where $\Delta$ is the conformal weight.
Through a conformal transformation the Hamiltonians on the plane and on the cylinder are related by
\begin{equation}
	L_0 + \bar L_0 - \frac{c}{12} = H_0 - \frac{1}{12},
\end{equation} 
(the last factor corresponds to a zero-point energy), $c$ being the central charge of Liouville theory
\begin{equation}
	c = 1 + 6 Q^2.
\end{equation} 
Comparing these equations with \eqref{mini:eq:schrodinger-mabuchi-p} teaches that the conformal dimension is related to $p$ by
\begin{equation}
	\Delta = \frac{Q^2}{4} + p^2.
\end{equation} 
Moreover the states $V_p = \e^{2i p \phi}$ on the cylinder are mapped to states $V_a = \e^{2a \phi}$ on the plane where the relation between $a$ and $p$ is
\begin{equation}
	a = \frac{Q}{2} + i p.
\end{equation} 
Since the Liouville theory is unitary the conformal weights should be positive, i.e.\ $\Delta \ge 0$, which implies $p \in \R_+$ or $p \in i [0, Q/2)$ (only half of the intervals are considered as a consequence of the reflection).

In the Liouville theory the generalization of the formula \eqref{mini:eq:3-point} computes the semi-classical approximation to the DOZZ structure constant $C(a_1, a_2, a_3)$ with the following weights~\cite{Schomerus:2003:RollingTachyonsLiouville}
\begin{equation}
	a_1 = \frac{Q}{2} + i p_1, \qquad
	a_2 = i p_2, \qquad
	a_3 = \frac{Q}{2} + i p_3.
\end{equation} 

\section{Complete Mabuchi Hamiltonian}
\label{app:adm}

In this section we consider the Mabuchi action where the Liouville field has been replaced using \eqref{quant:eq:relation-phi-K}
\begin{multline}
	\label{mab:action:mabuchi-K-with-bdy}
	S_M = \frac{\epsilon}{4\pi} \int \dd^2 \sigma\, \sqrt{g_0}\; \bigg[
			- \pi\chi \ell\, g_0^{\mu\nu} \pd_\mu K \pd_\nu K
			+ \left( \frac{4\pi \chi}{A_0} - R_0 \right) K \\
			+ \frac{2}{A_0 \ell} \left(1 + \frac{A_0 \ell}{2}\, \lap_0 K \right) \left(\ln \frac{A}{A_0} \left(1 + \frac{A_0 \ell}{2}\, \lap_0 K \right) - 1 \right)
		\bigg],
\end{multline} 
where $\epsilon = \pm 1$ is used to consider both Euclidean and Lorentzian signatures.
Moreover the parameter $\ell$ is used to consider both the unscaled and the scaled actions simultaneously: $\ell = 1$ corresponds to \eqref{quant:action:mabuchi} (with the boundary term) and $\ell = (\pi\chi)^{-1}$ to \eqref{quant:action:mabuchi-rescaled-bdy}.

The strategy for computing the Hamiltonian is to perform first an ADM decomposition~\cite{Arnowitt:2004:DynamicsGeneralRelativity} of the metric in order to extract the time derivative of the Kähler potential before using the Ostrogradski formalism~\cite{Woodard:2015:TheoremOstrogradsky} since the action is of second order in time.
Note that the background metric $g_0$ is fixed and for this reason its components are not dynamical.
In particular it is not necessary to decompose the curvature $R_0$ and to apply the full ADM formalism.

The ADM decomposition of the metric is
\begin{equation}
	g_{0\mu\nu} = \e^{2\rho}
		\begin{pmatrix}
			\epsilon N^2 + M^2 & M \\
			M & 1
		\end{pmatrix}, \qquad
	g_{0}^{\mu\nu} = \frac{\epsilon \e^{-2\rho}}{N^2}
		\begin{pmatrix}
			1 & - M \\
			- M & \epsilon N^2 + M^2
		\end{pmatrix}.
\end{equation} 
where $\rho$, $N$ and $M$ are functions of the coordinates (note that the matrix part is flat).
This decomposition is valid locally and the topology is hidden in the coordinates~\cite{Polchinski:2005:StringTheory-1} and in the values of the conformal factor $\rho$.
In particular the latter has to be singular if $\chi \neq 0$ since
\begin{equation}
	4\pi \chi = \int \dd^2 \sigma \sqrt{g_0}\, R_0
		= \int \dd^2 \sigma\, \pd^2 \rho
\end{equation} 
where $\pd^2$ is the flat Laplacian.
Nonetheless we will not work directly with its value and the fact that it contains singularities does not matter.

The squareroot of the metric determinant is
\begin{equation}
	\sqrt{\abs{g_0}} = N \e^{2\rho}.
\end{equation} 
The Laplacian is
\begin{multline}
	\label{mab:eq:laplacian-adm}
	\lap_0 = \frac{\epsilon\, \e^{-2\rho}}{N^2} \bigg[
			\pd^2_\tau
			+ (\epsilon N^2 + M^2)\, \pd^2_\sigma
			- 2 M \pd_\tau \pd_\sigma
			+ \left(\frac{M N'}{N} - M' - \frac{\dot N}{N} \right) \pd_\tau
			\\
			+ \left(2 (\epsilon N' N + M' M) + \frac{M \dot N}{N} - (\epsilon N^2 + M^2) \frac{N'}{N} - \dot M \right) \pd_\sigma
		\bigg].
\end{multline} 
The kinetic term of the action \eqref{mab:action:mabuchi-K-with-bdy} is
\begin{equation}
	- \frac{\epsilon \pi\chi \ell}{2}\, \sqrt{g_0} g_0^{\mu\nu} \pd_\mu K \pd_\nu K
		= - \frac{\pi\chi \ell}{2 N} \left( 
				\dot K^2
				- 2 M \dot K K'
				+ (\epsilon N^2 + M^2) K'^2
			\right).
\end{equation} 
It is not needed to decompose the curvature $R_0$ because only $K$ is dynamical, not the background metric $g_0$.

Now one can apply the Ostrogradski formalism.
The independent variables\footnotemark{} are $\{ K, \dot K \}$ with conjugate momenta $\{ \mc P, P \}$
\footnotetext{%
	One could also consider the $K'$ to be an independent variable.
	Then the last term in $\mc P$ would correspond to the derivative of its conjugate momentum.
	The resulting Hamiltonian is then equivalent to the one obtained below upon integration by part.
}%
\begin{equation}
	P = \frac{\pd L}{\pd \ddot K}, \qquad
	\mc P = \frac{\pd L}{\pd \dot K} - \pd_\tau P - \pd_\sigma \frac{\pd L}{\pd \dot K'}
\end{equation} 
and the Hamiltonian reads
\begin{equation}
	H = \mc P \dot K + P \ddot K - L
\end{equation} 
where the Lagrangian is normalized such that
\begin{equation}
	S = \frac{1}{2\pi} \int \dd^2 \sigma\, L.
\end{equation} 

The momentum $P$ is
\begin{equation}
	P = \frac{1}{2 N}\, \ln \frac{A}{A_0} \left(1 + \frac{A_0 \ell}{2}\, \lap_0 K \right).
\end{equation} 
Using the relation \eqref{quant:eq:relation-phi-K-rescaled} one recognizes that the RHS of $P$ is proportional to $\phi$ and for this reason one can perform a canonical transformation to invert the roles of position and momentum (after having computed the Hamiltonian)
\begin{equation}
	N P = \phi, \qquad
	\dot K = - N \Pi.
\end{equation} 
Moreover the above expression can be used to solve for $\ddot K$ in terms of the canonical variables using \eqref{mab:eq:laplacian-adm}.
The second momentum $\mc P$ is
\begin{equation}
	\mc P = \pi \chi \ell\, \Pi
		- \frac{1}{N}\, \dot \phi
		+ \frac{2 M}{N}\, \phi'
		+ \pi \chi \ell\, \frac{M}{N}\, K'
		- \frac{1}{N} \left( \frac{M N'}{N} - M' \right) \phi.
\end{equation} 
Ultimately one finds the Hamiltonian
\begin{equation}
	\label{mab:eq:hamiltonian-mabuchi-adm}
	\begin{aligned}
		H =\ & \frac{\pi\chi \ell}{2}\, N\, \Pi^2
			- N\, \Pi \mc P
			+ 2 M\, \Pi \phi'
			+ \left( M' - \frac{M N'}{N} - \frac{\dot N}{N} \right) \Pi \phi
			+ \pi\chi \ell\, M\, \Pi K'
			\\
			&+ \frac{\pi\chi \ell}{2}\, (\epsilon N^2 + M^2) K'^2
			- \frac{1}{N}\, (\epsilon N^2 + M^2) K'' \phi
			\\
			&+ \frac{1}{N} \left( \dot M + (\epsilon N^2 + M^2) \frac{N'}{N} - 2 (\epsilon N N' + M M') - \frac{M \dot N}{N} \right) K' \phi
			\\
			&- \frac{\epsilon N}{2}\, \e^{2\rho} \left( \frac{4\pi \chi}{A_0} - R_0 \right) K
			+ \frac{\epsilon N}{A \ell}\, \e^{2\rho} \e^{2\phi}
			- \frac{2 \epsilon N}{A_0 \ell}\, \e^{2\rho} \phi.
	\end{aligned}
\end{equation} 

Several limits can be taken from this Hamiltonian (from now on $\epsilon = -1$).
In particular in the flat gauge
\begin{equation}
	N = 1, \qquad
	M = 0, \qquad
	\rho = 0,
\end{equation} 
but keeping $R_0, \chi \neq 0$ for comparison, one finds
\begin{equation}
		H = \frac{\pi\chi \ell}{2}\, \Pi^2
			- \Pi \mc P
			+ \frac{1}{2} \left( \frac{4\pi \chi}{A_0} - R_0 \right) K
			+ \frac{2}{A_0 \ell}\, \phi
			- \frac{1}{A \ell}\, \e^{2\phi}
\end{equation} 
in the absence of spatial dependence.
For $\ell = 1$ this Hamiltonian has no particularly meaningful limit $R_0, \chi \longrightarrow 0$, and in particular it contains a term linear in $\phi$ which could lead to an instability.
On the other hand for $\ell = (\pi\chi)^{-1}$ one recovers \eqref{eq:ostrogradski:hamiltonian-mabuchi-phi}, and from there one can recover \eqref{mini:eq:hamiltonian-mabuchi} in the limit $R_0, \chi \longrightarrow 0$.

\printbibliography[heading=bibintoc]

\end{document}